# Topological bulk solitons in a nonlinear photonic Chern insulator

Rujiang Li,[1,*] Xiangyu Kong,[1] Dongkai Hang,[1] Guoyi Li,[1] Hongyu Hu,[1] Hao Zhou,[2] Yongtao Jia,[1] Pengfei Li,[3,4] and Ying Liu[1,†]

[1]Key Laboratory of Antennas and Microwave Technology, School of Electronic Engineering, Xidian University, Xi'an 710071, China

[2]Science and Technology on Antenna and Microwave Laboratory, Nanjing Research Institute of Electronic Technology, Nanjing 210013, China

[3]Department of Physics, Taiyuan Normal University, Jinzhong 030619, China

[4]Institute of Computational and Applied Physics, Taiyuan Normal University, Jinzhong 030619, China

[*]rujiangli@xidian.edu.cn

[†]liuying@mail.xidian.edu.cn

## ABSTRACT

Nonlinearities in lattices with topological band structures can induce topological interfaces in the bulk of structures and give rise to bulk solitons in the topological bandgaps. Here we study a photonic Chern insulator with saturable nonlinearity and show the existence of topological bulk solitons. The fundamental bulk solitons exhibit as semi-vortex solitons, where only one pseudospin component has a nonzero vorticity. The bulk solitons have equal angular momentum at different valleys. This phenomenon is a direct outcome of the topology of the linear host lattice and the angular momentum can be changed by switching the sign of the nonlinearity. The bulk solitons bifurcate from the linear bulk band edge and terminate when their powers saturate. We find that these bulk solitons are stable within the whole spectrum range. Moreover, these bulk solitons are robust against lattice disorders both from on-site energies and hopping amplitudes. Our work extends the study of Chern insulators into the nonlinear regime and highlights the interplay between topology and nonlinearity.

# INTRODUCTION

Topology is a branch of mathematics which is concerned with integer-valued quantities that are preserved under continuous deformations. Different objects are topologically equivalent as long as they have the same topological invariant. By describing the global structure of the wave function in momentum space over the Brillouin zone (BZ) as a topological invariant, the concept of topology was introduced into condensed matter physics with the advent of topological insulators, which are insulating in the bulk but conducting on their surfaces even in the presence of impurities [1, 2]. Shortly after the discovery of topological insulators, the photonic analogues of topological insulators based on quantum Hall effect, namely the photonic Chern insulators were proposed and realized [3-5]. Later, based on the photonic Chern insulators which are characterized by the integer-valued Chern numbers, photonic Floquet topological insulators [6, 7], photonic spin-Hall insulators [8-12] and photonic valley-Hall insulators [13-20] were reported successively, and the field of topological photonics has flourished [21-24]. Among all these two-dimensional (2D) photonic topological insulators, photonic Chern insulators are the most reliable designs to date that provide truly unidirectional and backscattering-immune wave transport due to the genuine breakdown of time-reversal symmetry [25]. They have been reported experimentally in gyromagnetic photonic crystals at microwave frequencies [5] and active resonator networks at optical frequencies [26].

The investigation of photonic topological insulators has been extended into the nonlinear regime with the inclusion of the interparticle interactions [27]. By combining topology with nonlinearity, a kind of solitons dubbed "bulk solitons" are found (without the linear counterpart) in the bulk topological bandgap [28]. Unlike the nonlinear edge states or edge solitons [29–36], the bulk solitons reside in the bulk of materials. Meanwhile, these bulk solitons are also different to the conventional lattice solitons [37]. Lattice solitons are formed as a result of the balance between coupling (commonly referred as diffraction) and nonlinearity. Bulk solitons occur at the nonlinearity induced topological interfaces. Until now, 1D bulk solitons have been demonstrated in Su-Schrieffer-Heeger (SSH) chains [38–41], and 2D bulk solitons have been demonstrated in both photonic Floquet insulators [28, 42, 43] and photonic valley-Hall insulators [38, 44].

In this paper, we extend the photonic Chern insulators into the nonlinear regime and study the bulk solitons under a saturable nonlinearity, since the properties of static Chern insulators are fundamentally different from that of driven Floquet insulators [7, 45]. In contrast to the photonic

Floquet insulators which are driven lattice systems with time-reversal symmetry breaking replaced by *z*-reversal symmetry breaking [28, 42], our photonic Chern insulator is a non-driven lattice and the actual time-reversal symmetry is broken. We find that the bulk solitons bifurcate from the linear bulk band edge of the energy spectrum (rather than the quasi-energy spectrum of a photonic Floquet topological insulator [28, 42]), and terminate when their powers saturate. The fundamental bulk solitons exhibit as semi-vortex solitons, where only one pseudospin component has a nonzero vorticity. Due to the topology of the photonic Chern insulator in the linear limit, the bulk solitons have equal angular momentum at different valleys. This phenomenon is a direct outcome of the topology of the linear host lattice. The angular momentum of the bulk soliton is changed when the self-focusing nonlinearity is switched to the self-defocusing nonlinearity. Finally, we also demonstrate the stability and robustness of the bulk solitons. Our results extend the study of Chern insulators into the nonlinear regime and highlight the interplay between topology and nonlinearity.

**RESULTS**

**Model**

We study a photonic Chern insulator based on the celebrated Haldane model. It is described by a tight-binding Hamiltonian with a real nearest-neighbor (NN) hopping (black lines), a complex next-nearest-neighbor (NNN) hopping (red and blue arrows) and on-site staggering energies (white and black circles), as shown in Fig. 1(a). The wave functions at sublattice sites A (white circles) and B (black circles) can be treated as two pseudospins. The length of the nearest-neighbor bonds is $a_0$ and the lattice period is $a = \sqrt{3}a_0$. Two reciprocal lattice vectors are $\mathbf{b}_1 = \left(2\pi/a, 2\pi/\sqrt{3}a\right)$ and $\mathbf{b}_2 = \left(0, 4\pi/\sqrt{3}a\right)$, where $\mathbf{b}_i \cdot \mathbf{a}_j = 2\pi\delta_{ij}$. In the Brillouin zone (BZ) shown in Fig. 1(b), K and K′ are the two valleys.

We define two sets of vectors $\mathbf{e}_{1,2,3}$ and $\mathbf{v}_{1,2,3}$ for the NN hopping and NNN hopping, respectively. Based on the Hamiltonian for the Haldane lattice, we add the saturable nonlinear terms to the on-site energies. Then the coupled equations for the two pseudospin components $\psi_{A,B}$ are

$$\begin{aligned} \mathrm{i}\frac{\partial \psi_A(\mathbf{r},t)}{\partial t} = (\omega_A + N_A)\psi_A(\mathbf{r},t) + t_1 \sum_{i=1,2,3} \psi_B(\mathbf{r}+\mathbf{e}_i,t) \\ + t_2 e^{\mathrm{i}\phi} \sum_{i=1,2,3} \psi_A(\mathbf{r}+\mathbf{v}_i,t) + t_2 e^{-\mathrm{i}\phi} \sum_{i=1,2,3} \psi_A(\mathbf{r}-\mathbf{v}_i,t), \end{aligned} \quad (1)$$

$$i\frac{\partial \psi_B(\mathbf{r},t)}{\partial t} = (\omega_B + N_B)\psi_B(\mathbf{r},t) + t_1 \sum_{i=1,2,3} \psi_A(\mathbf{r}-\mathbf{e}_i,t)$$
$$+ t_2 e^{i\phi} \sum_{i=1,2,3} \psi_B(\mathbf{r}-\mathbf{v}_i,t) + t_2 e^{-i\phi} \sum_{i=1,2,3} \psi_B(\mathbf{r}+\mathbf{v}_i,t), \quad (2)$$

where $t_1$ is the real NN hopping, $t_2 e^{\pm i\phi}$ are the complex NNN hoppling, $\omega_{A,B}$ are the on-site energies, and $N_{A,B} = g\frac{|\psi_{A,B}(\mathbf{r},t)|^2}{1+\sigma|\psi_{A,B}(\mathbf{r},t)|^2}$ characterizes the saturable nonlinearity with the nonlinear parameter $g$ and saturation coefficient $\sigma$. Eqs. (1)-(2) are the discrete nonlinear Schrödinger equations with two components. In the absence of optical losses, the normalized power $P = \sum_{\mathbf{r}}(|\psi_A(\mathbf{r})|^2 + |\psi_B(\mathbf{r})|^2)$ is conserved.

For simplicity, we assume that $t_2 > 0$ and $\phi = \pi/2$, since we are focusing on a photonic Chern insulator. In the linear limit with $g = 0$, due to the time-reversal symmetry breaking, the bulk bands split and lead to a complete bandgap with K and K′ valleys (See Supplementary Note 1). Specifically, the bulk bands are characterized by Chern numbers, implying the presence of chiral edge modes. In the nonlinear regime, we are interested in the modes near the valleys, where the bands have linear dispersions. We substitute $\psi_{A,B}(\mathbf{r},t) = \psi_{1,2}(\mathbf{r},t)\exp(i\mathbf{K}_\pm \cdot \mathbf{r})$ into Eqs. (1)-(2), where $\mathbf{K}_\pm$ correspond to K and K′ valleys, respectively. Expanding the wave functions around $\mathbf{r}$ and neglecting the higher-order terms, we get

$$i\frac{\partial \psi_1}{\partial t} = (d_0 + d_3 + N_1)\psi_1 \pm iv_F \frac{\partial \psi_2}{\partial x} + v_F \frac{\partial \psi_2}{\partial y}, \quad (3)$$

$$i\frac{\partial \psi_2}{\partial t} = (d_0 - d_3 + N_2)\psi_2 \pm iv_F \frac{\partial \psi_1}{\partial x} - v_F \frac{\partial \psi_1}{\partial y}, \quad (4)$$

where $d_0 = \frac{\omega_A + \omega_B}{2}$ is the frequency shift, $d_3 = \frac{\omega_A - \omega_B}{2} \pm 3\sqrt{3}t_2$ is the effective mass that opens the linear bulk bandgap, $v_F = \frac{\sqrt{3}}{2}t_1 a$ is the group velocity, and $N_{1,2} = g\frac{|\psi_{1,2}|^2}{1+\sigma|\psi_{1,2}|^2}$. These equations are valid as long as $\left|\frac{\partial \psi_{1,2}(\mathbf{r})}{\partial \mathbf{r}} \cdot \mathbf{e}_{1,2,3}\right| \ll |\psi_{1,2}(\mathbf{r})|$ and $\left|\frac{\partial \psi_{1,2}(\mathbf{r})}{\partial \mathbf{r}} \cdot \mathbf{v}_{1,2,3}\right| \ll |\psi_{1,2}(\mathbf{r})|$ which require that $\psi_{A,B}$ should be smoothly distributed in the lattice and the length of the nearest-neighbor bonds should be small enough [38]. Note that Eqs. (3)-(4) reduce to the continuous linear Dirac equations when $d_3 = g = 0$ [46]. Besides, unlike the 2D nonlinear Dirac equations where nonlinearity is equivalent to a mass term [44], in our model nonlinearity changes both the effective frequency shift and effective mass (See Supplementary Note 2).

The solutions for bulk solitons can be found by solving Eqs. (3)-(4) using Newton-conjugate-

gradient method [47]. With the harmonic time dependence $e^{-i\omega t}$, we have

$$\mathbf{L}_0 \boldsymbol{\Psi} = 0, \tag{5}$$

where

$$\mathbf{L}_0 = \begin{pmatrix} d_0 + d_3 - \omega + N_1 & v_F\left(\pm i\dfrac{\partial}{\partial x} + \dfrac{\partial}{\partial y}\right) \\ v_F\left(\pm i\dfrac{\partial}{\partial x} - \dfrac{\partial}{\partial y}\right) & d_0 - d_3 - \omega + N_2 \end{pmatrix} \tag{6}$$

and $\boldsymbol{\Psi} = (\psi_1, \psi_2)^T$. We iteratively update the solution as $\boldsymbol{\Psi}_{n+1} = \boldsymbol{\Psi}_n + \Delta\boldsymbol{\Psi}_n$, and the updated amount $\Delta\boldsymbol{\Psi}_n$ is computed from the linear Newton-correction equation

$$\mathbf{L}_1^\dagger \mathbf{L}_1 \Delta\boldsymbol{\Psi}_n = -\mathbf{L}_1^\dagger \mathbf{L}_0 \boldsymbol{\Psi}_n, \tag{7}$$

where

$$\mathbf{L}_1 = \begin{pmatrix} d_0 + d_3 - \omega + gN_1' & v_F\left(\pm i\dfrac{\partial}{\partial x} + \dfrac{\partial}{\partial y}\right) \\ v_F\left(\pm i\dfrac{\partial}{\partial x} - \dfrac{\partial}{\partial y}\right) & d_0 - d_3 - \omega + gN_2' \end{pmatrix}, \tag{8}$$

$N_{1,2}' = \dfrac{2|\psi_{1,2}|^2 + \psi_{1,2}^2 C}{1 + \sigma|\psi_{1,2}|^2} - \sigma|\psi_{1,2}|^2 \psi_{1,2} \dfrac{\psi_{1,2}^* + \psi_{1,2} C}{\left(1 + \sigma|\psi_{1,2}|^2\right)^2}$, and $C$ is the complex conjugate operator.

**Bulk solitons**

For simplicity, we let $\omega_A = \omega_B = 10$, and thus we have $d_3 = \pm 3\sqrt{3} t_2$ for K and K' valleys, respectively. The other parameters are $a = 1$, $t_1 = 2/\sqrt{3}$, $t_2 = 1/3\sqrt{3}$, and $\sigma = 1$. In the linear limit, the bulk bands create a bandgap where the central frequency is $\omega = d_0 = 10$ and the bandgap half-width is $|d_3| = 1$. In the nonlinear case, we only study the fundamental modes because the higher-order solitons are usually unstable [48]. Fig. 2 shows the bulk solitons for the self-focusing nonlinearity with $g = 1$ and the frequency is $\omega = 9.1$. For the bulk soliton at K valley [Figs. 2(a)-(b)], the pseudospin component $\psi_2$ decreases monotonously from a finite value to zero along the radial direction, but the amplitude of $\psi_1$ features a hump at a nonzero radius. This single hump behavior with a ring shape is different with the soliton profile for a nonlinear Dirac equation [44].

From the phase distributions shown in the insets, the phase of the pseudospin component $\psi_1$ increases clockwise by $2\pi$ and thus has a vorticity with $l_1 = -1$ around $r = 0$. The component $\psi_2$ has a zero vorticity with $l_2 = 0$. The vorticity is defined as $l_{1,2} = \frac{1}{2\pi}\oint_L \nabla[\arg(\psi_{1,2})] \cdot d\mathbf{l}$ with the time dependence dropped [49]. Since only one component has a nonzero vorticity, the bulk soliton at K valley is a semi-vortex soliton.

The semi-vortex soliton is not a direct outcome of the nonlinearity-induced localization of the linear bulk states. The linear bulk states exhibit sublattice polarizations at the K and K′ valleys, and have nonzero vorticities induced by the phase terms $\exp(i\mathbf{K}\cdot\mathbf{r})$ and $\exp(i\mathbf{K}'\cdot\mathbf{r})$ (See Supplementary Note 1). While for the bulk soliton here, it exhibits the vorticity in the absence of the phase terms. Similar to the bulk solitons in photonic Floquet insulators [28, 42] and photonic valley-Hall insulators [38, 44], our bulk soliton in the photonic Chern insulator can also be understood as the topological edge state residing in the bulk. Nonlinearity induces a mass inversion and creates a topological domain wall in the bulk (See Supplementary Note 2). Along the domain wall, the localized state known as the Jackiw-Rebbi Dirac boundary mode exists [50]. The bulk soliton shown in Figs. 2(a)-(b) can be directly obtained from the Jackiw-Rebbi mode by smoothly changing the parameters $\sigma$ and $\omega$.

We also would like to note that the bulk soliton is topologically protected from two aspects. First, the bulk soliton is protected by the topology of the linear host lattice in momentum space. The nonzero Chern number of the linear bulk band implies the existence of a bulk topological bandgap. As long as the linear bulk bandgap is open, nonlinearity can create a topological domain wall and support the existence of the bulk soliton. Second, the bulk soliton is also topologically protected in real space because one of its components has nonzero vorticity.

Similarly, from Figs. 2(c)-(d), the bulk soliton at K′ is also a semi-vortex soliton. The difference is that, the pseudospin component $\psi_1$ has a zero vorticity with $l_1 = 0$ and the component $\psi_2$ has a nonzero vorticity with $l_2 = -1$. Such behavior is related to the topology of the linear host lattice. In the linear Haldane model, time-reversal symmetry is broken but inversion symmetry is preserved. Since $d_3$ has opposite signs at the K and K′ valleys, we have equal Berry curvatures with

$\Omega(K) = \Omega(K')$ and non-zero Chern numbers for the bulk bands [25]. Due to the inversion symmetry with $d_3(K) = -d_3(K')$, according to Eqs. (3)-(4), if we make transformations $\psi_1 \to -\psi_2$ and $\psi_2 \to \psi_1$ to the equations at K valley, we can get the equations at K' valley. Thus, the bulk solitons at K and K' valleys both rotate clockwise with a phase difference of π. In other words, the bulk solitons at different valleys have equal angular momentum. Note that for a photonic valley-Hall insulator, the bulk solitons at K and K' valleys rotate in opposite directions (See Supplementary Note 3). Fig. 3 shows the bulk solitons for the self-defocusing nonlinearity with $g = -1$ and the frequency is $\omega = 10.9$. The bulk solitons at K and K' valleys are also semi-vortex solitons and they rotate counterclockwise with a phase difference of π. Compared with the bulk solitons in Fig. 2, the rotating direction is switched by the sign of nonlinearity. This implies that the self-defocusing nonlinearity can also be used to induce the topological interface and create bulk solitons in photonic topological insulators.

To give more insights into the bulk solitons, we transform Eqs. (3)-(4) into polar coordinate system and get

$$i\frac{\partial \psi_1}{\partial t} = v_F e^{\mp i\varphi}\left(\pm i\frac{\partial}{\partial r} + \frac{1}{r}\frac{\partial}{\partial \varphi}\right)\psi_2 + (d_3 + N_1)\psi_1, \qquad (9)$$

$$i\frac{\partial \psi_2}{\partial t} = v_F e^{\pm i\varphi}\left(\pm i\frac{\partial}{\partial r} - \frac{1}{r}\frac{\partial}{\partial \varphi}\right)\psi_1 - (d_3 - N_2)\psi_2. \qquad (10)$$

Here the term $d_0$ is dropped by a gauge transformation $\psi_{1,2}(\mathbf{r},t) = \psi_{1,2}(\mathbf{r},t)\exp(-id_0 t)$. We seek for 2D bulk solitons with harmonic time dependence and radial symmetry. First, we study the equations at K valley. If $d_3 > 0$, we substitute

$$\begin{pmatrix} \psi_1 \\ \psi_2 \end{pmatrix} = \begin{pmatrix} iue^{-i(m+1)\varphi} \\ ve^{-im\varphi} \end{pmatrix} e^{-i\omega t} \qquad (11)$$

into Eqs. (9)-(10) and get the following equations

$$v_F\left(\frac{d}{dr} + \frac{m+1}{r}\right)u + (d_3 + \omega - N(v))v = 0, \qquad (12)$$

$$v_F\left(\frac{d}{dr} - \frac{m}{r}\right)v + (d_3 - \omega + N(u))u = 0, \qquad (13)$$

where $N(u,v) = g\frac{|u,v|^2}{1+\sigma|u,v|^2}$. These equations can be numerically solved and admit the existence

of solutions for semi-vortex solitons [38]. If $d_3 < 0$, we substitute

$$\begin{pmatrix} \psi_1 \\ \psi_2 \end{pmatrix} = \begin{pmatrix} ue^{im\varphi} \\ ive^{i(m+1)\varphi} \end{pmatrix} e^{-i\omega t} \tag{14}$$

into Eqs. (9)-(10) and get the following equations

$$v_F \left( \frac{d}{dr} - \frac{m}{r} \right) u + \left( -d_3 - \omega + N(v) \right) v = 0, \tag{15}$$

$$v_F \left( \frac{d}{dr} + \frac{m+1}{r} \right) v + \left( -d_3 + \omega - N(u) \right) u = 0. \tag{16}$$

These equations are equivalent to Eqs. (12)-(13) and also have semi-vortex soliton solutions. Then we study the equations at K′ valley. If $d_3 > 0$, we substitute

$$\begin{pmatrix} \psi_1 \\ \psi_2 \end{pmatrix} = \begin{pmatrix} -iue^{i(m+1)\varphi} \\ ve^{im\varphi} \end{pmatrix} e^{-i\omega t} \tag{17}$$

and get the same equations as Eqs. (12)-(13). If $d_3 < 0$, we substitute

$$\begin{pmatrix} \psi_1 \\ \psi_2 \end{pmatrix} = \begin{pmatrix} ue^{-im\varphi} \\ -ive^{-i(m+1)\varphi} \end{pmatrix} e^{-i\omega t} \tag{18}$$

and get the same equations as Eqs. (15)-(16).

In our model, we have $d_3 > 0$ and $d_3 < 0$ at K and K′ valleys, respectively. For the bulk solitons at K valley, the solutions with $m = 0$ and $m = -1$ in Eq. (11) are the fundamental modes for self-focusing nonlinearity with $g > 0$ and self-defocusing nonlinearity with $g < 0$, respectively. The solution for the bulk soliton in Figs. 2(a)-(b) is $(\psi_1, \psi_2)^T = (iue^{-i\varphi}, v)^T$, and the bulk soliton in Figs. 3(a)-(b) can be written as $(\psi_1, \psi_2)^T = (u, -ive^{i\varphi})^T$. Similarly, for the bulk solitons at K′ valley, the solutions with $m = 0$ and $m = -1$ in Eq. (18) are the fundamental modes for self-focusing nonlinearity with $g > 0$ and self-defocusing nonlinearity with $g < 0$, respectively. The solution for the bulk soliton in Figs. 2(c)-(d) is $(\psi_1, \psi_2)^T = (u, -ive^{-i\varphi})^T$, and the bulk soliton in Figs. 3(c)-(d) can be written as $(\psi_1, \psi_2)^T = (iue^{i\varphi}, v)^T$. Thus, we validate the fact that the bulk solitons at different valleys have the equal angular momentum and the rotating direction can be changed by the sign of nonlinearity. Similar discussion can also be carried out for the bulk solitons in a photonic valley-Hall insulator (See Supplementary Note 3).

We discuss the mode distributions of the bulk solitons. As an example, we consider the bulk solitons at K valley under the self-focusing nonlinearity [Figs. 2(a)-(b)] and Eqs. (12)-(13) with $m = 0$ are their governing equations. In the limit of $r \to 0$, we have $u(r=0) = 0$ and $v'(r=0) = 0$. In the limit of $r = \infty$, mode localization requires that $u(r=\infty) = v(r=\infty) = 0$. The contribution from the nonlinear terms is small and Eqs. (12)-(13) are reduced to the linear differential equations. Since the bulk solitons reside in the bulk bandgap with $|\omega| < |d_3|$ [51], the solutions are

$$u(r) = CK_{m+1}\left(\frac{\sqrt{d_3^2 - \omega^2}}{v_F} r\right), \tag{19}$$

$$v(r) = C\sqrt{\frac{d_3 - \omega}{d_3 + \omega}} K_m\left(\frac{\sqrt{d_3^2 - \omega^2}}{v_F} r\right), \tag{20}$$

with $m = 0$. Considering these limiting values, the pseudospin component $\psi_1$ exhibits a hump at a nonzero radius and the component $\psi_2$ decreases monotonously from a finite value to zero along the radial direction. These results agree well with the soliton distribution in Figs. 2(a)-(b).

**Existence**

We study the existence of the bulk solitons. We only show the results for the bulk solitons at K valley since the results at K′ valley can be obtained from the transformations $\psi_1 \to -\psi_2$ and $\psi_2 \to \psi_1$. Figs. 4(a)-(b) show the radial distributions of the bulk solitons, where the parameters are $a = 1$, $\omega_A = \omega_B = 10$, $t_1 = 2/\sqrt{3}$, $t_2 = 1/3\sqrt{3}$, $g = 1$, and $\sigma = 1$. Under the self-focusing nonlinearity, the amplitudes of both the two components increase when the frequency $\omega$ approaches the central frequency of the bandgap. We define $P_{1,2} = \int |\psi_{1,2}(\mathbf{r})|^2 d^2r$ as the optical powers for the pseudospin components $\psi_{1,2}$. From Fig. 4(c), both the curves for $P_1$ and $P_2$ are monotonic. The bulk solitons bifurcate from the lower band edge at $\omega = 9$ and reside in the topological bandgap. When the optical powers are large enough, the bulk soliton terminates and its frequency saturates near $\omega = 10$, which corresponds to the center of the bandgap. To quantify the degree of localization of the bulk solitons as a function of frequency, we plot the inverse participation ratios (IPRs) which are defined as $\text{IPR}_{1,2} = \frac{\int |\psi_{1,2}(\mathbf{r})|^4 d^2r}{\left(\int |\psi_{1,2}(\mathbf{r})|^2 d^2r\right)^2}$, as a measure of the localizations of the bulk solitons. From Fig.

4(d), the size (i.e. spatial extent) of the bulk solitons decreases to a minimum near the center of the existence range. In other words, bulk solitons closer to the linear band edge and bandgap center have larger spatial extents. This quantitative result agrees with the amplitude distributions in Figs. 4(a)-(b). We also show the power and IPR of the bulk solitons under the self-defocusing nonlinearity with $g = -1$ in Figs. 4(e) and (f), respectively. In contrast to the bulk solitons under the self-focusing nonlinearity, the bulk solitons under the self-defocusing nonlinearity bifurcate from the upper band edge at $\omega = 11$. Thus, the sign of nonlinearity can also change the bifurcation and existence range of the bulk solitons.

**Stability**

We study the stability of the bulk solitons. As an example, we only show the results for the bulk solitons at K valley under the self-focusing nonlinearity, where the parameters are $a = 1$, $\omega_A = \omega_B = 10$, $t_1 = 2/\sqrt{3}$, $t_2 = 1/3\sqrt{3}$, $g = 1$, and $\sigma = 1$. First, we perform the linear stability analysis by following a standard linearization procedure. The solution is sought at the frequency $\delta$ in the form of

$$\psi_1 = e^{-i\omega t}\left(\phi_1 + \varepsilon_1 e^{-i\delta t} + \mu_1^* e^{i\delta^* t}\right), \tag{21}$$

$$\psi_2 = e^{-i\omega t}\left(\phi_2 + \varepsilon_2 e^{-i\delta t} + \mu_2^* e^{i\delta^* t}\right), \tag{22}$$

where $\phi_{1,2} e^{-i\omega t}$ are the unperturbed soliton solution, $\varepsilon_{1,2}$ and $\mu_{1,2}$ are the infinitesimal amplitudes of the perturbations. Obviously, the bulk solitons are linearly stable if $\delta$ is real. They are linearly unstable if the imaginary part of $\delta$, namely the growth rate, is positive. Substituting the perturbed solutions into Eqs. (3)-(4), we get the linearized equations regarding to $\varepsilon_1$, $\mu_1$, $\varepsilon_2$, and $\mu_2$. The linearized equations can then be numerically solved using Fourier collocation method combined with Newton-conjugate-gradient method [47].

Fig. 5(a) shows the growth rates Im($\delta$) for the bulk solitons at different frequencies. In the whole frequency range, the growth rates are in the order of $10^{-6}$. Note that the perturbations may come from both the amplitudes and phases. For the perturbation eigenmode with a certain vorticity $q$, the solution can be written as

$$\begin{pmatrix} \psi_1 \\ \psi_2 \end{pmatrix} = \left[ \begin{pmatrix} \tilde{\phi}_1 \\ \tilde{\phi}_2 \end{pmatrix} + \begin{pmatrix} \tilde{\varepsilon}_1 \\ \tilde{\varepsilon}_2 \end{pmatrix} e^{-iq\varphi} e^{-i\delta t} + \begin{pmatrix} \tilde{\mu}_1^* \\ \tilde{\mu}_2^* \end{pmatrix} e^{iq\varphi} e^{i\delta^* t} \right] \begin{pmatrix} e^{-i\varphi} \\ 1 \end{pmatrix} e^{-i\omega t}. \qquad (23)$$

We show the perturbation eigenmodes $\varepsilon_{1,2}$ with $\omega = 9.35$ in Figs. 5(b)-(c), since the growth rate is largest at this frequency [corresponding to the red dot in Fig. 5(a)]. From the figures, the growth rates shown in Fig. 5(a) correspond to the perturbation eigenmodes with $q = 0$. The contributions from the higher-order perturbations with $q \neq 0$ are negligible. Based on the above results and using the Vakhitov-Kolokolov (VK) stability criterion, we can conclude that the bulk solitons are linearly stable within the whole spectrum range, although the growth rates are not exactly equal to zero due to the numerical errors. VK criterion predicts the instability of the solitons with $dP/d\omega < 0$ and provides the necessary stability condition $dP/d\omega > 0$ [52, 53]. Usually for solitons without vortices, $dP/d\omega > 0$ is also a sufficient stability condition, because the perturbation eigenmodes have $q = 0$ and there are no azimuthal perturbations with $q \neq 0$ [54]. From Fig. 4(c), the power $P_2$ dominates the total power and we have $dP_2/d\omega > 0$ within the whole spectrum range. Since the pseudospin component $\psi_2$ has a zero vorticity, the contributions from the azimuthal perturbations with $q \neq 0$ (namely the perturbations from phases) are negligible. Thus, according to the VK criterion, the bulk solitons are linearly stable. Physically, the bulk solitons are stabilized by the saturable nonlinearity. The bulk solitons under the nonsaturable Kerr nonlinearity become unstable when the frequency exceeds a threshold, because there exists a range for $dP_2/d\omega < 0$ (See Supplementary Note 3). The saturable nonlinearity suppresses the continuous decrease of the spatial size of the solitons and changes the slopes of the existence curves. Then the high-frequency solitons become stable and all the bulk solitons in the whole spectrum range are stable.

Next, we further confirm the stability of the bulk solitons by directly simulating Eqs. (3)-(4) based on Runge-Kutta method in the spectrum domain. The evolution time is $t = 10000$, which is long enough to observe the soliton dynamics. When the initial input is selected as the soliton solution with $\omega = 9.35$, we observe the stationary evolution of the bulk soliton, as shown in Figs. 6(a)-(b). When ±5% noises with uniform distributions are added to the initial input, although the bulk soliton experiences a transverse displacement, the bulk soliton still exhibits as a semi-vortex soliton and the radii of the isosurfaces are invariant during the temporal evolution [Figs. 6(c)-(d)], unlike the breathing or collapse of the unstable solitons under Kerr nonlinearity (See Supplementary Note 4). This behavior confirms that, although the pseudospin component $\psi_1$ has a nonzero vorticity, the bulk solitons are only disturbed by the radially symmetric perturbations with $q = 0$ and there is no radial symmetry breaking. These results valid the fact that the bulk solitons are stable in the whole spectrum range.

**Robustness**

Since the bulk solitons are nonlinearity induced topological edge states, it is crucial to study their robustness against lattice disorders. We study the robustness of the bulk solitons by adding disorders to the parameters in Eqs. (3)-(4) and observing the temporal evolution of the bulk solitons under a noiseless input. As an example, we only show the robustness of the bulk soliton with $\omega$ = 9.35 at K valley under the self-focusing nonlinearity. The parameters are $a = 1$, $\omega_A = \omega_B = 10$, $t_1 = 2/\sqrt{3}$, $t_2 = 1/3\sqrt{3}$, $g = 1$, and $\sigma = 1$. The evolution time is also set as $t = 10000$. The parameters $\omega_{A,B}$, $t_1$, $t_2$, $g$, and $\sigma$ can be treated as two kinds: $\omega_{A,B}$, $g$, and $\sigma$ are the (equivalent) on-site energies; and $t_1$ and $t_2$ are the hopping amplitudes.

First, we add ±5% disorders with uniform distributions to both $g$ and $\sigma$, and the result is shown in Figs. 7(a)-(b). Since the nonlinearity is a perturbation to the on-site energies, disorders from the nonlinear parameter $g$ and saturation coefficient $\sigma$ only induce a small drift of the soliton center in *xy* plane [see the soliton trajectory in Fig. 7(a) and Supplementary Movie 1] and the soliton profile has no deformation [Fig. 7(b)]. Here we only demonstrate the amplitude of the pseudospin component $\psi_1$ at the output with $t = 10000$ because $\psi_1$ has a small optical power [Fig. 4(c)] and it should be more sensitive to the disorders.

Second, we add ±5% disorders to both $\omega_A$ and $\omega_B$. In contrast to the first case, the bulk soliton shows a considerable random drift [Fig. 7(c) and Supplementary Movie 2]. Since periodic boundary conditions are imposed in the numerical simulations of Eqs. (3)-(4), the bulk soliton can move out the simulation domain from one boundary and reenter from the opposite boundary. Besides, from the amplitude of $\psi_2$ shown in Fig. 7(d), there is a deformation of the soliton profile. Considering the structure disorders, we find that the upper and lower band edges are governed by

$\omega_0 \mp \delta\omega \pm \sqrt{(-\delta\omega+1)^2}$, where $\omega_A = \omega_B = \omega_0$ and $\delta\omega$ is the strength of the disorders. Since the robustness of a topological edge state is guaranteed only when the disorders do not close the bulk bandgap, the disorder strength $\delta\omega$ should be smaller than 0.5. Thus, the robustness of bulk soliton is degraded due to the large disorders from on-site energies.

Third, we add ±5% disorders to $t_1$. Since the change of the NN hopping amplitude does not affect the central frequency and bandgap width of the linear bulk band structure, the bulk soliton in this case has no deformation [Fig. 4(f)]. During temporal evolution, the bulk soliton exhibits a random transverse drift because of the breakdown of the periodicity of the lattice [Fig. 4(e) and Supplementary Movie 3].

Finally, we add ±5% disorders to $t_2$. Although the NNN hopping controls the bandgap width, the bandgap exists as long as we have $t_2 \pm \delta t_2 \neq 0$. This condition is satisfied obviously. The bulk soliton does not deform during temporal evolution [Fig. 4(h)], although it also experiences a transverse drift [Fig. 4(g) and Supplementary Movie 4].

Regarding to the lattice disorders, our bulk solitons are robust to the disorders from $g$, $\sigma$, $t_1$, and $t_2$; and only weakly perturbed by the disorders from the on-site energies $\omega_{A,B}$. Experimentally it is possible to reduce the on-site disorders to less than ±5% or observe the bulk solitons within a short period [42]. Besides, the bulk solitons are also topologically protected in real space. From Figs. 7(b), (d), (f), and (h) and Supplementary Movies 1-4, although there are disorders from on-site energies or hopping amplitudes, the first pseudospin components still have nonzero vorticity. Thus, the bulk solitons are robust against the lattice disorders both from on-site energies and hopping amplitudes within the whole spectrum range, and the intervalley conversion of the bulk solitons is prohibited.

**Comparison with discrete bulk solitons**

In Figs. 2-7, the bulk solitons are obtained by solving the continuous model based on Eqs. (3)-(4). Here we show the discrete bulk solitons directly calculated from Eqs. (1)-(2) and compare them with the continuous ones. For simplicity, we only consider the discrete bulk solitons at K valley under the self-focusing nonlinearity. The parameters are $\omega_A = \omega_B = 10$, $a = 1$, $t_1 = 2/\sqrt{3}$,

$t_2 = 1/3\sqrt{3}$, $g = 1$, and $\sigma = 1$.

Figs. 8(a)-(b) show the discrete bulk soliton with the frequency $\omega = 9.1$. Similar to the continuous bulk soliton in Figs 2(a)-(b), the amplitude of $\psi_1$ features a hump at a nonzero radius and the pseudospin component $\psi_2$ exhibits a peak at the lattice center. From the phase distributions, near the lattice centers, the phase of the pseudospin component $\psi_1$ increases clockwise by $2\pi$ and the phase of the component $\psi_2$ is zero. The phase distortions away from the lattice centers are created by the discrete lattice configuration. Thus, the discrete bulk solitons are also semi-vortex solitons. Similarly, Figs. 8(c)-(d) show the discrete bulk soliton with the frequency $\omega = 9.9$. The discrete bulk soliton extends to more lattice sites and this behavior agrees with that of the continuous bulk solitons [Figs.4(a)-(b)]. Besides, in Figs. 8(e)-(f), we show the full lattice distribution of the discrete bulk soliton with $\omega = 9.9$, where (e) and (f) correspond to the amplitude and phase, respectively. Unlike the continuous bulk solitons [Figs.5 and 6], the discrete bulk solitons are unstable within the whole spectrum range (See Supplementary Note 5). To stabilize the discrete bulk solitons, we need to decrease the lattice period $a$ and increase the NN hopping $t_1$ [51].

## DISCUSSION

We have several remarks here. First, from the above results, the (continuous) bulk solitons reside in a complete topological bandgap, and they are stable and robust within the whole spectrum range. Because of these features, it is feasible to observe the bulk solitons experimentally in a photonic Chern insulator. In the microwave frequency, linear photonic Chern insulators have been realized in gyromagnetic photonic crystals [5, 55]. Nonlinearity can be introduced into the photonic crystals using nonlinear elements such as varactors [56]. In the optical frequency, a linear photonic Chern insulator is also realized using active resonator networks [26]. Nonlinearity can be easily incorporated by using nonlinear materials and high input power. Besides, electrical circuit lattices have been proposed recently as the low-frequency counterparts of the photonic topological insulators [57]. Since both Chern circuits and nonlinear topological circuits have been demonstrated [33, 58], it is feasible to implement a nonlinear Chern circuit and observe the bulk solitons. The theoretical results presented here are not specific to classical photonic systems. They are also applicable to photonic lattices realized in atomic systems [59–61] and exciton-polariton systems [40,

62, 63].

Second, we would like to compare our results with the bulk solitons in a photonic Floquet insulator [28]. Our paper reveals several features that were not discovered previously. Our bulk solitons are found under the saturable nonlinearity, in contrast to the Kerr nonlinearity used in the photonic Floquet insulator. Due to the saturable nonlinearity, our bulk solitons are stable and robust, but the bulk solitons in the photonic Floquet insulator are unstable. We also find that the angular momenta of the bulk solitons at different valleys are linked to the topology of the linear host lattice and the angular momenta can be switched by the sign of the nonlinearity. These behaviors were not reported in previous publications.

Third, we would like to point out that the bulk solitons also exist at other k points. At a general k point, Eqs. 3-(4) are generalized to the general form of the nonlinear Dirac equations, which admit the existence of soliton solutions (See Supplementary Note 6). However, in order to get a more accurate result when compared with the discrete bulk solitons calculated from Eqs. (1)-(2), higher-order expansions should be considered because the dispersions are nonlinear at general k points.

## CONCLUSIONS

In conclusion, our work predicts the existence of topological bulk solitons in a nonlinear photonic Chern insulator and we find that the bulk solitons exhibit as semivortex solitons. Our bulk solitons at different valleys have the equal angular momentum and the angular momentum can be switched by the sign of the nonlinearity. Moreover, the bulk solitons are stable and robust within the whole spectrum range. Our work can be extended to study bulk solitons in other photonic topological insulators, such as the photonic spin-Hall insulators [8–12].

## METHODS

For the continuous model [Eqs. (3)-(4)], Newton-conjugate-gradient method is used to seek for the bulk soliton solutions. The stability of the bulk solitons is studied using the linear stability analysis by following a standard linearization procedure and the linearized equations are numerically solved using Fourier collocation method combined with Newton-conjugate-gradient method. The temporal evolution of the bulk solitons is studied using the Runge-Kutta method in the spectrum domain. For the discrete model [Eqs. (1)-(2)], the Newton's method is used to seek for the solutions for the discrete bulk solitons. The stability of the discrete bulk solitons is also studied using the standard linear stability analysis.

## DATA AVAILABILITY

The data to produce Figs. 2-8 and Supplementary Movies 1-4 are available from the corresponding author upon reasonable request.

## CODE AVAILABILITY

The program code of this study is available from the corresponding author upon reasonable request.

## ACKNOWLEDGMENTS

R.L. and X.K. were sponsored by the National Natural Science Foundation of China (NSFC) under Grant No. 12104353. D.H., G.L., and H.H are supported by the Fundamental Research Funds for the Central Universities. P.L. was sponsored by the National Natural Science Foundation of China (NSFC) (11805141), the Applied Basic Research Program of Shanxi Province (201901D211424), and the Scientific and Technological Innovation Programs of Higher Education Institutions in Shanxi (STIP) (2021L401). Y.L. was sponsored by the National Natural Science Foundation of China (NSFC) under Grant No. 61871309 and the 111 Project. The numerical calculations in this paper were supported by High-Performance Computing Platform of Xidian University. The authors also acknowledge the Beijing Super Cloud Computing Center (BSCC) for providing HPC resources that have contributed to part of the research results reported within this paper. URL: http://www.blsc.cn/.

## AUTHOR CONTRIBUTIONS

R.L. conceived the idea. R.L., X.K., and P.L. performed the analytical calculations. R.L., X.K., D.H., G.L., H.H., H.Z., Y.J. and P.L. performed the numerical calculations. R.L. wrote the manuscript. R.L. and Y.L. supervised the project.

## COMPETING INTERESTS

The authors declare no competing interests.

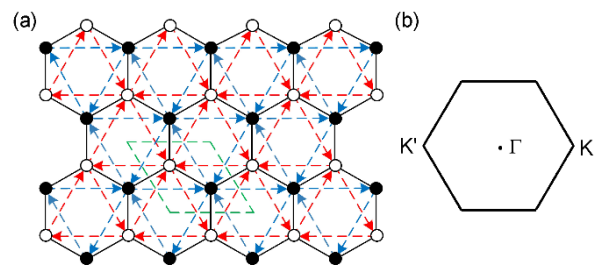

FIG. 1. **Schematic and Brillouin zone (BZ) of the Haldane model.** (a) Schematic of the Haldane model with a real nearest-neighbor (NN) hopping (black lines), a complex next-nearest-neighbor (NNN) hopping (red and blue arrows), and on-site staggering energies (white and black circles). The rhombus with green lines is the unit cell. (b) BZ of the Haldane model, where K and K′ are the two valleys.

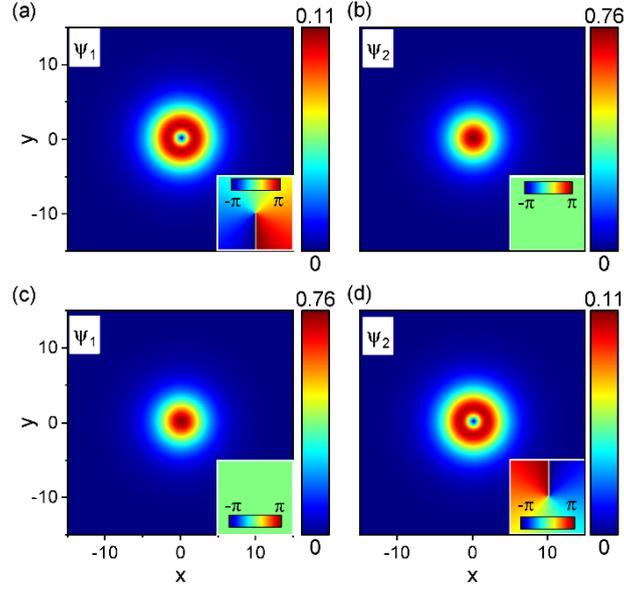

FIG. 2. **Bulk solitons under the self-focusing nonlinearity with *g* = 1.** (a)-(b) The amplitudes of the two pseudospin components $\psi_{1,2}$ of the bulk soliton at K valley. (c)-(d) The amplitudes of the two pseudospin components of the bulk soliton at K′ valley. The color bars in (a)-(d) denote the amplitudes and the insets in (a)-(d) are the phases. The frequencies of the bulk solitons are $\omega = 9.1$.

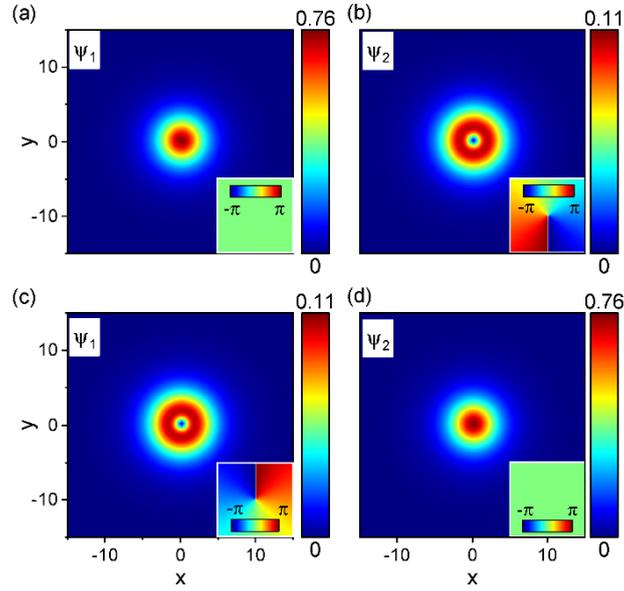

FIG. 3. **Bulk solitons under the self-defocusing nonlinearity with *g* = −1.** (a)-(b) The amplitudes of the two pseudospin components $\psi_{1,2}$ of the bulk soliton at K valley. (c)-(d) The amplitudes of the two pseudospin components of the bulk soliton at K′ valley. The color bars in (a)-(d) denote the amplitudes and the insets in (a)-(d) are the phases. The frequencies of the bulk solitons are $\omega = 10.9$.

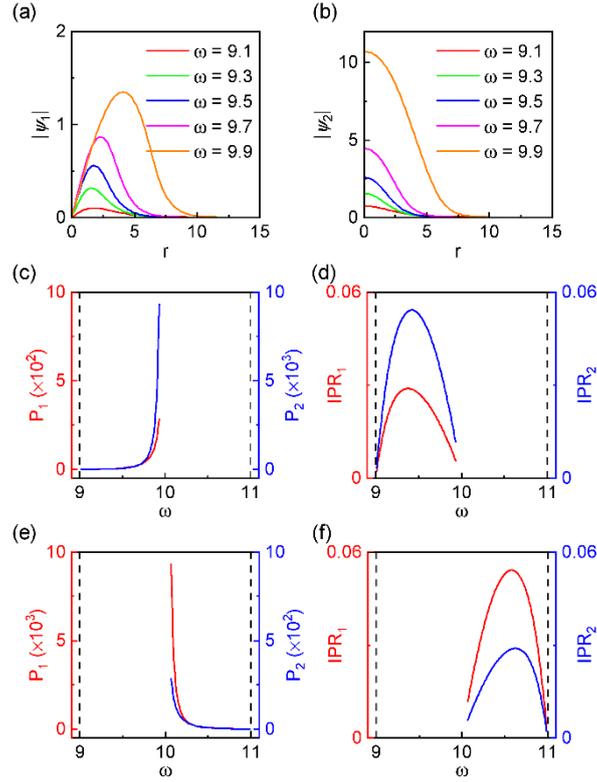

FIG. 4. **Radial distributions and existence curves of the bulk solitons.** (a)-(b) The radial distributions of the bulk solitons under the self-focusing nonlinearity with $g = 1$. (c)-(d) Power [(c)] and inverse participation ratio (IPR) [(d)] of the bulk soliton under the self-focusing nonlinearity with $g = 1$. (e)-(f) Power [(e)] and IPR [(f)] of the bulk soliton under the self-defocusing nonlinearity with $g = -1$. The dashed lines in (b)-(f) denote the lower and upper band edges.

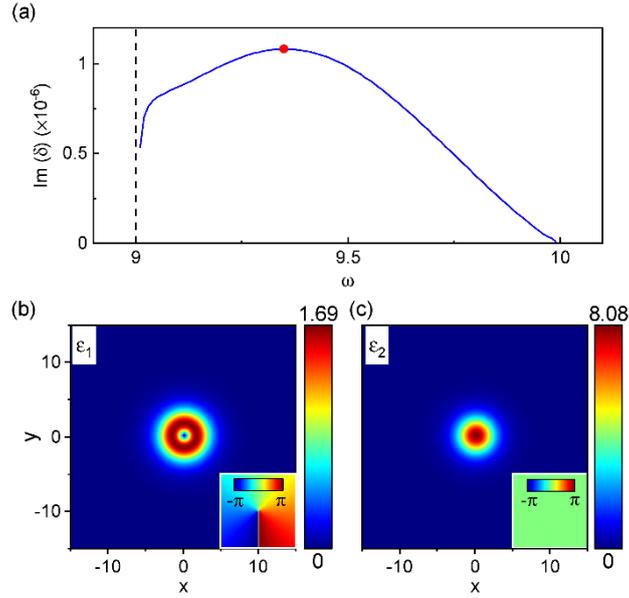

FIG. 5. **Stability of the bulk solitons.** (a) Growth rates Im($\delta$) for the bulk solitons at different frequencies. The dashed line denotes the lower band edge. (b)-(c) The perturbation eigenmodes $\varepsilon_{1,2}$ with the frequency $\omega = 9.35$, which correspond to the red dot in (a). The color bars in (b)-(c) denote the amplitudes of the perturbation eigenmodes.

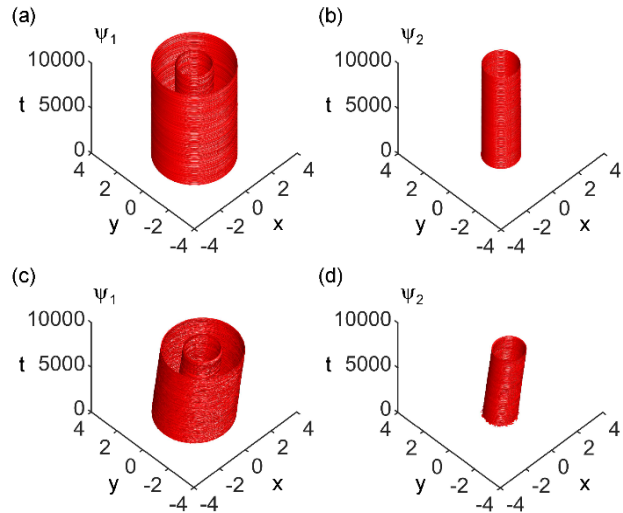

FIG. 6. **Temporal evolutions of the bulk solitons.** Isosurfaces for the temporal evolutions of the bulk solitons without [(a)-(b)] and with [(c)-(d)] noises added to the initial input. The soliton frequencies are $\omega = 9.35$.

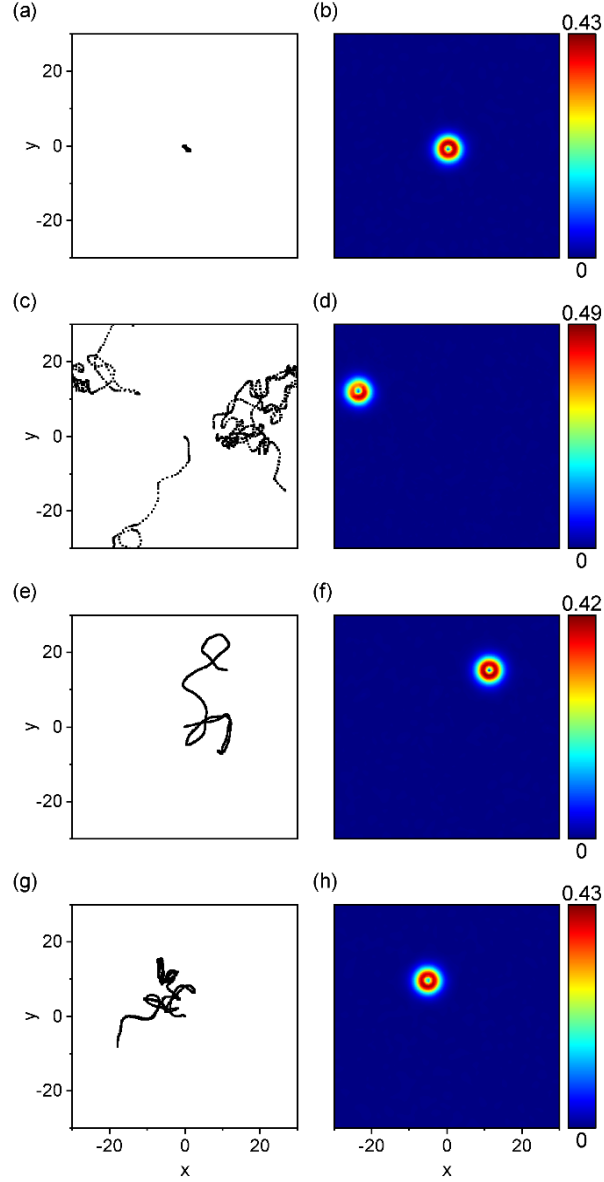

FIG. 7. **Robustness of the bulk solitons.** Robustness of the bulk soliton with the frequency $\omega$ = 9.35 at K valley under the self-focusing nonlinearity, when ±5% disorders are added to the nonlinear parameter $g$ and saturation coefficient $\sigma$ [(a)-(b)], on-site energies $\omega_{A,B}$ [(c)-(d)], nearest-neighbor (NN) hopping $t_1$ [(e)-(f)], and next-nearest-neighbor (NNN) hopping $t_2$ [(g)-(h)], respectively. (a), (c), (e), and (g) show the trajectories of the soliton centers under the temporal evolution; and (b), (d), (f), and (h) show the amplitudes of the pseudospin component $\psi_1$ at the output with $t$ = 10000. The color bars in (b), (d), (f), and (h) denote the amplitudes.

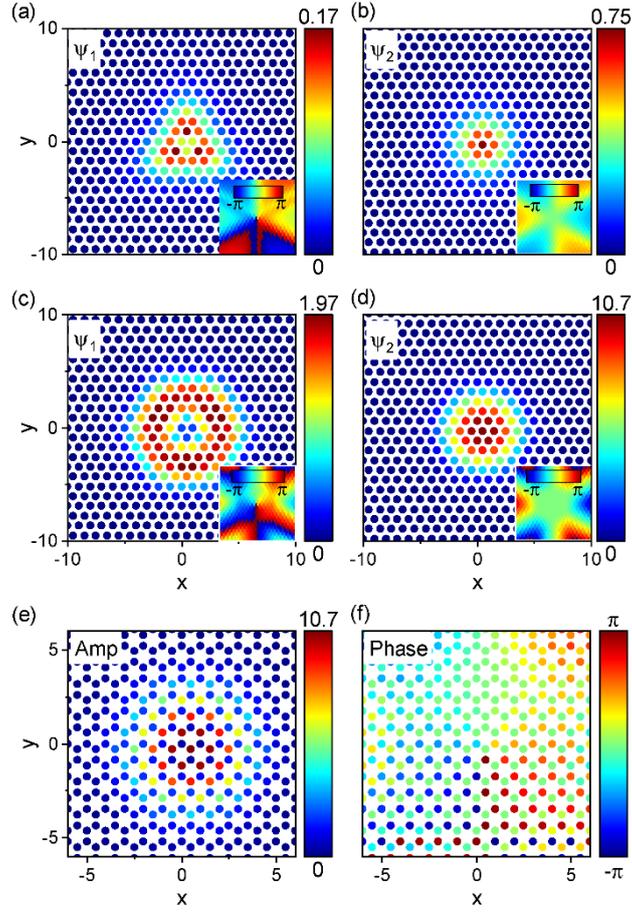

FIG. 8. **Discrete bulk solitons.** Discrete bulk solitons at K valley under the self-focusing nonlinearity with $g = 1$. (a)-(b) The amplitudes of the two pseudospin components $\psi_{1,2}$ of the discrete bulk soliton with the frequency $\omega = 9.1$. (c)-(d) The amplitudes of the two pseudospin components $\psi_{1,2}$ of the discrete bulk soliton with the frequency $\omega = 9.9$. The color bars in (a)-(d) denote the amplitudes and the insets in (a)-(d) are the phases. (e)-(f) The full lattice distribution of the discrete bulk soliton with the frequency $\omega = 9.9$, where the color bars in (e) and (f) correspond to the amplitude and phase, respectively.